\documentclass[aps,prb,twocolumn,groupedaddress,floatfix]{revtex4-1}
\usepackage{amssymb,amsmath}
\usepackage{graphicx}
\usepackage{subfigure}
\usepackage[english]{babel}
\usepackage{float}
\usepackage{color}

\begin{document}
\newcommand{\be}{\begin{equation}}
\newcommand{\ee}{\end{equation}}
\newcommand{\rojo}[1]{\textcolor{red}{#1}}

\title{Seltrapping in flat band lattices with nonlinear disorder}

\author{Danilo Rivas and Mario I. Molina}
\affiliation{Departamento de F\'{\i}sica, Facultad de Ciencias, Universidad de Chile, Casilla 653, Santiago, Chile}

\date{\today }

\begin{abstract} 
We study the selftrapping properties of an initially localized excitation in several flat band lattices, in the presence of nonlinear (Kerr) disorder. In the weak nonlinearity regime, the dynamics is controlled by the degeneracy of the bands leading to a linear form of selftrapping. In the strong nonlinearity regime, the dynamics of the excitations depends strongly on the local environment around the initial excitation site that leads to a highly fluctuating selfrapping profile. It is shown that the spreading of the wavefunction in the presence of random nonlinearity, is always completely inhibited about fifty percent of all cases. The role of degeneracy associated with the flatbands is explained with the help of the $n$-simplex system, which can be solved in closed form.

\end{abstract}

\maketitle

\section{Introduction}
The phenomenon of selftrapping of excitations in nonlinear lattices has been an active research field for many years, from the time it was realized that some solutions   of the coupled electron-phonon problem in biomolecules, featured stable,  propagating localized excitations, termed discrete solitons. At the time, it provided a possible way to understand the propagation of excitations along some molecules\cite{davidov,davidov2}. In the semiclassical approach to the problem, the vibrational coordinates are treated classically while the electron is treated quantum mechanically. Further, if the vibrational degrees of freedom are assumed to be enslaved to the electronic ones, one arrives to an effective electronic equation known as the Nonlinear Discrete Schr\"{o}dinger (DNLS) equation\cite{DNLS,eilbeck,johansson}[see Eq.(1)].

In addition to the condensed matter or biophysical context, the DNLS has also appeared in other physical contexts, such as coupled waveguide arrays in optics\cite{optics1,optics2} and Bose-Einstein condensates 
in coupled magneto-optical traps\cite{BEC1,BEC2}. One of the striking features of the DNLS equation (see below) is that it leads to discrete solitons: the localization of the excitation in a small region of the lattice and capable of propagating for relatively long distances along a quasi onedimensional chain, with little dispersion.

Along the years the DNLS has been examined for several lattices and various dimensionalities, and under a variety of conditions ranging from disorder in the on-site energies\cite{flach,aubry,molina_disorder},
effect of nonlinear impurities\cite{NLimpurity1,NLimpurity2,NLimpurity3,NLimpurity4,NLimpurity5,NLimpurity6}, nonlocal effects\cite{nonlocal}, disorder of the nonlinear 
parameter\cite{DisorderChi}, and long-range effects\cite{LongRange}.
Perhaps the most striking feature observed is the robustness of the discrete soliton, under many of these  conditions. Its existence can be argued based on the local character of the DNLS (Eq.1): Starting with a nonlinear lattice and assuming
 that a localized excitation exists, say of size $d$, then the nonlinearity is only appreciable near the soliton position, and the rest of the lattice can be taken as linear, in a first approximation.  We are left with a sort of nonlinear impurity of size $d$ and, since we know a breaking of translational invariance gives rise to a localized mode, we should have a localized excitation. 
This closes the self-consistent argument. There remains finer details like the minimum value of $\chi$ for trapping to occur. 

Another mechanism for generating localized modes this time in linear systems has gained recent interest: Flat bands. Simply stated, a flat band system is a periodic system whose spectra contains flat bands. The presence of a flat band implies the existence of a set of entirely degenerate states, which do not display evolution in time. In an optical context they are interesting since they allow the long-distance propagation without distortion of shapes based on combinations of these flat modes. These states rely on a precise geometrical interference condition, and have been studied and observed in optical and photonic lattices\cite{flat1,flat2,flat3,flat4}, graphene\cite{graphene1,graphene2}, superconductors\cite{sc1,sc2,sc3,sc4}, fractional quantum Hall systems\cite{hall1,hall2,hall3}, and exciton-polariton condensates\cite{condensate1,condensate2}.

In this work we examine the interplay between degeneracy and nonlinear disorder in several lattices that possess flat bands. In particular, we will focus on the trapping dynamics of these excitations and study their thresholds for selftrapping. We will show that, al small nonlinearities, degeneracy effects takes place  inducing a linear localization effect, while at high and disordered nonlinearities, selftrapping is strongly dependent on the local vicinity of the initial site, giving rise to a highly fluctuating selftrapping profile. The transport properties are also affected by the existence of flat bands and we will show that there are conditions under which nonlinear disorder does not affect the system transport.

\section{The model}
The Nolinear Schr\"{o}dinger (DNLS) equation is given by
\be
i {d C_{{\bf n}}\over{d z}}+\epsilon_{{\bf n}} C_{{\bf n}} + \sum_{{\bf m}} V_{{\bf n},{\bf m}} C_{{\bf m}} + \chi_{{\bf n}} |C_{{\bf n}}|^2 C_{{\bf n}}=0,\label{1}
\ee
where $C_{{\bf n}}$ is the electronic or optical excitation amplitude at site ${\bf n}$, 
$\epsilon_{{\bf n}}$ is a site energy term, $V_{{\bf n},{\bf m}}$ is the linear coupling between sites ${\bf n}$ and ${\bf m}$. Finally, $\chi_{{\bf n}}$ is the nonlinear parameter at site ${\bf n}$. In a condensed matter context, it corresponds to the square of the electron-phonon coupling at site ${\bf n}$. 
\begin{figure}[t]
 \includegraphics[width=\linewidth]{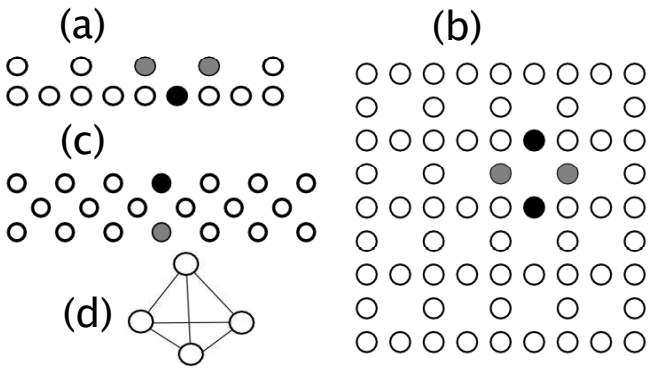}
  \caption{Schematic view of a stub (a), Lieb (b) and diamond (c) lattice. The simplex regular array for D=3 is shown in (d). Fundamental flatband modes for each lattice are shown in black and gray. Sites in black (gray) possess amplitudes 
  $A (-A)$ or viceversa.}
  \label{fig1}
\end{figure}
In this work we focus on the special case $\epsilon_{{\bf n}}=0$ and 
$\sum_{{\bf m}} V_{{\bf n},{\bf m}}=V$ for ${\bf n,m}$ nearest neighbors, zero otherwise. The nonlinear parameter
will be taken from a bivalued distribution: $\chi_{{\bf n}}=\chi\ \  \mbox{or}\ \  0$ with fifty-fifty percent probability. Initially, we place the excitation on a single site (${\bf n}=0$) of the lattice (or on a given waveguide in an optical waveguide array). Later on, we will excite the flat band mode and examine its time evolution under nonlinear disorder.

Figure 1 shows the flat band lattices to be considered here: Stub (a), Lieb (b), diamond(c) and a regular $n-simplex$. The last one is not a lattice, but an array of $n$ equidistant sites in dimension $D=n+1$ and is almost completely degenerate, as we will show below. For $D=0$ we have a single site ($n=1$), for $D=1$ we have a dimer ($n=2$), for $D=2$ we have an equilateral triangle ($n=3$), for $D=3$ we have a regular tetrahedron ($n=4$), 
\begin{figure}[t]
  \includegraphics[scale=0.45]{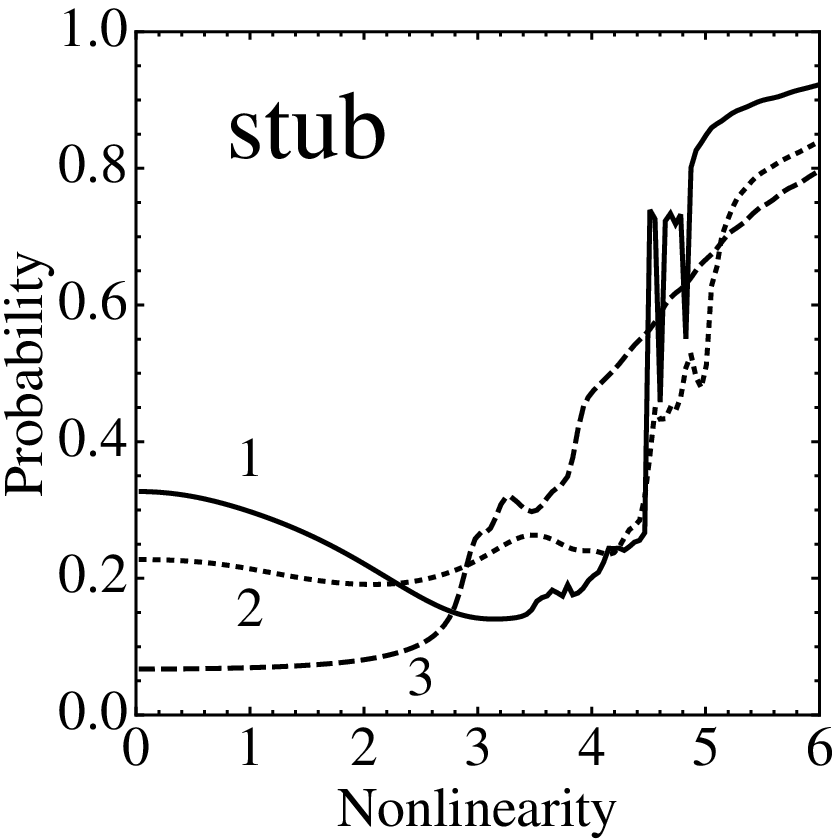}
 \hspace{0.1cm}
 \includegraphics[scale=0.45]{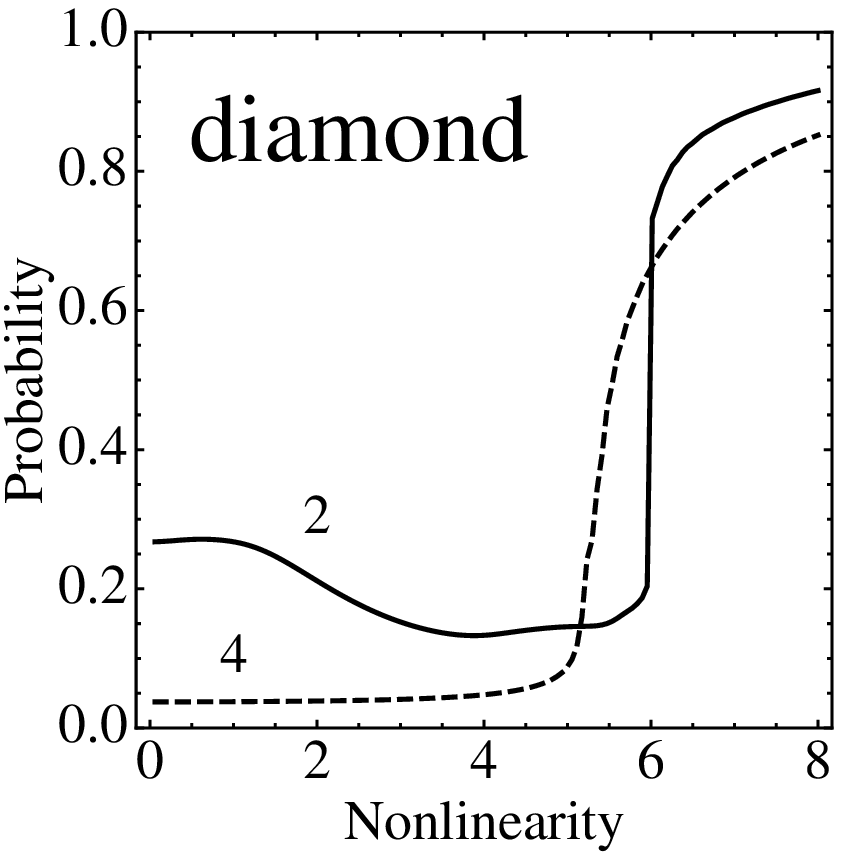}
 \includegraphics[scale=0.45]{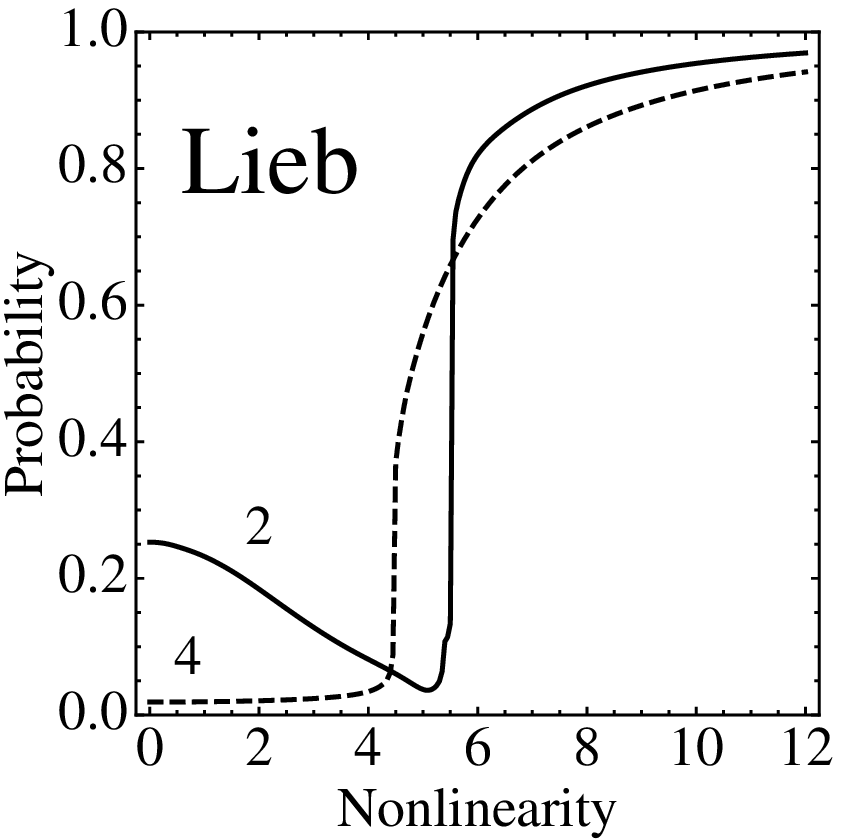}
 \hspace{0.1cm}
 \includegraphics[scale=0.45]{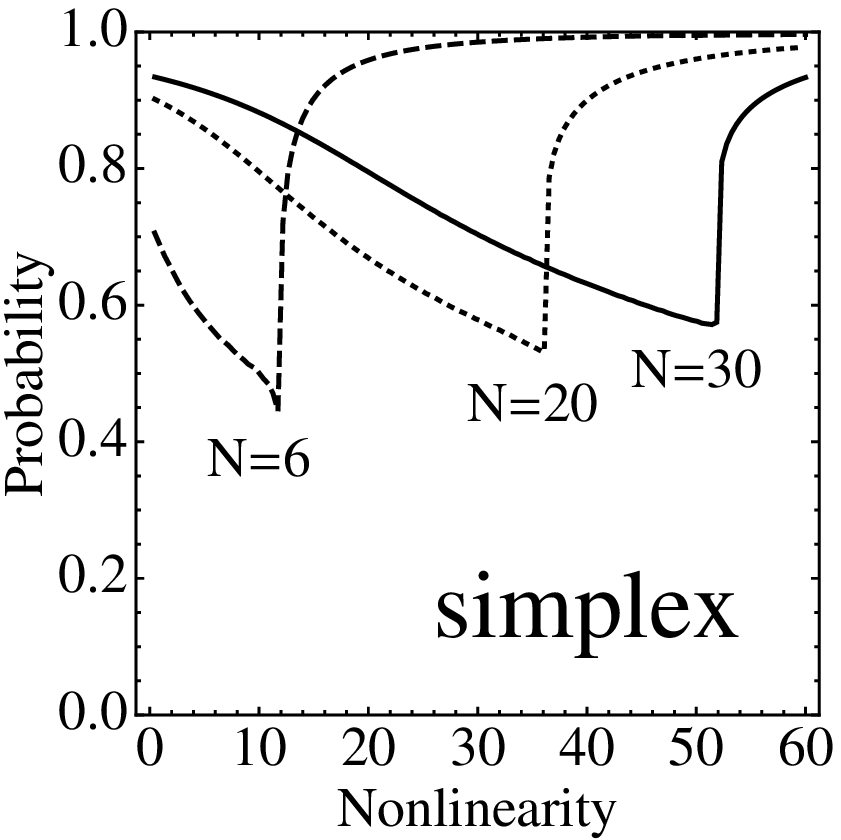}
  \caption{Time-averaged probability of finding the excitation at the initial site vs the nonlinearity parameter, for different lattices. The numerical labels for each curve denote the number of nearest neighbors of the initial site. For the {\em simplex} array, we plot the corresponding curves for different {\em simplex} sizes.}
\label{fig2}
\end{figure}
and so on. For each lattice we have also indicated the form of its fundamental flatband mode. For instance, for the Lieb lattice, this mode is a kind of ring $A,-A,A,-A$
which is characterized for having no transversal time evolution due to complete phase cancellation\cite{dany}.This is also true for the stub and diamond lattices. We will see that, when the initial site for the dynamical evolution of an observable overlaps partially with the flatband mode, it gives rise to partial trapping at small nonlinearity.

In order to ascertain the trapping of the excitation, 
we define the time-average probability at the initial site as
\be
\langle P_{0} \rangle = \lim_{T\rightarrow \infty} {1\over{T}} \int_{0}^{T} |C_{0}(t)|^2 dt.
\ee
To measure the spreading of the excitation as a function of time, we use the root mean square (RMS) displacement
\be
\sigma^{2}(t) = {\sum_{{\bf n}} ({\bf n}-{\bf n}_{0})^2 |C_{{\bf n}}(t)|^2 \over{\sum_{{\bf n}} |C_{{\bf n}}(t)|^2}},
\ee
and the participation ratio
\be
R(t)= { ( \sum_{{\bf n}} |C_{{\bf n}}(t)|^2 )^2  \over{\sum_{{\bf n}}|C_{{\bf n}}(t)|^4}}.
\ee
For a completely extended mode, $R = O(N)$, while for a completely localized state, $R = O(1)$.

\section{Uniform nonlinearity}
\label{uniform}
We start by examining $\langle P_{0}\rangle$ in the absence of   random nonlinearity $\chi_{\bf n}=\chi$. Results are shown in Fig.2, where the several curves denote the various inequivalent initial excitation sites that are possible for each lattice.
In all cases we notice that there are nonlinearity values 
at which abrupt selftrapping begins. This selftrapping transition is roughly similar for all these lattices, and its presence is in consonance with previous studies in nonlinear lattices\cite{nl}. However, there is a new ingredient here: At small nonlinearity strengths, selftrapping can be different for different initial sites of a given lattice. At some sites it is close to zero, while at others it is finite, indicating some degree of {\bf linear} selftrapping. For the simplex case, the only parameter left is the number of sites
of this generalized ``tetrahedron''. Here, we also notice that, in addition 
\begin{figure}[t]
 \includegraphics[scale=0.45]{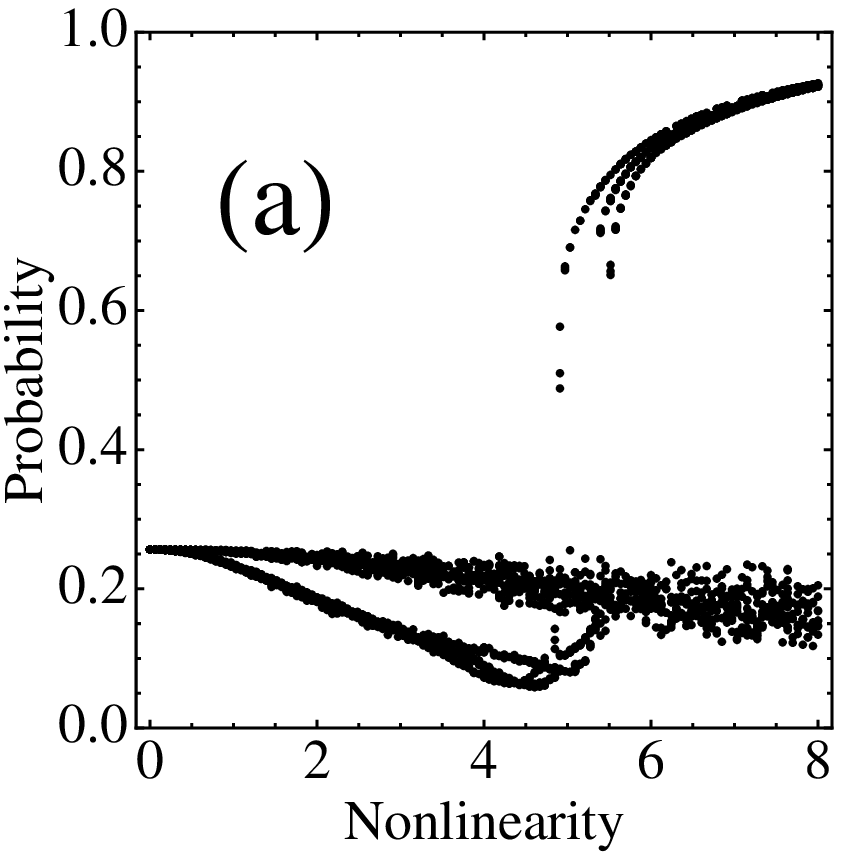}
 \hspace{0.1cm}
 \includegraphics[scale=0.45]{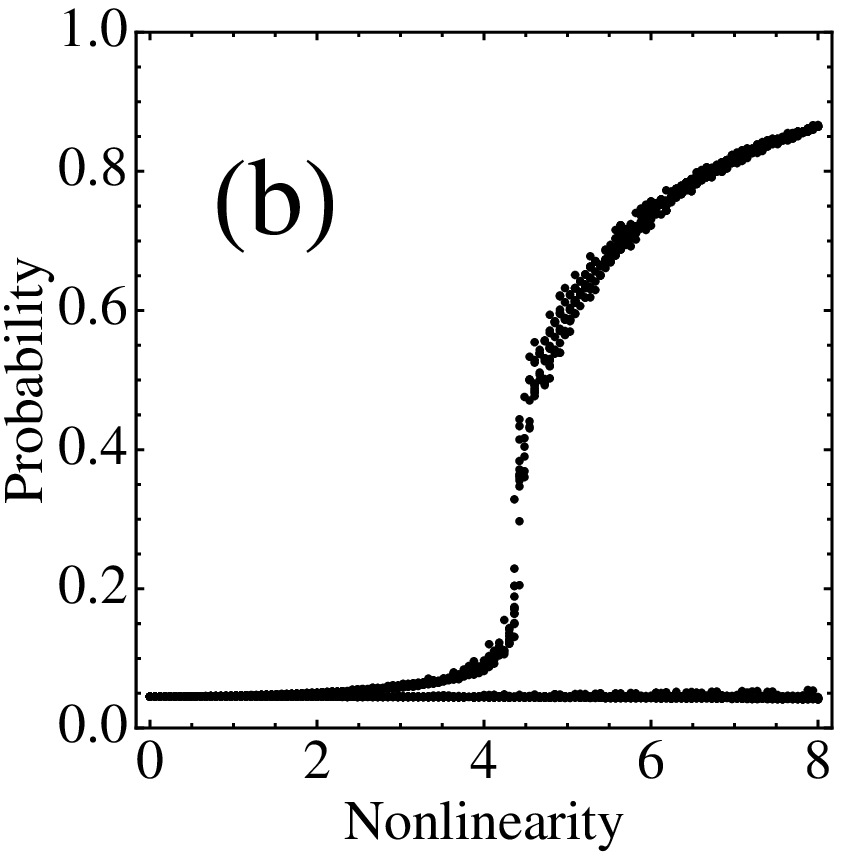}
 \includegraphics[scale=0.45]{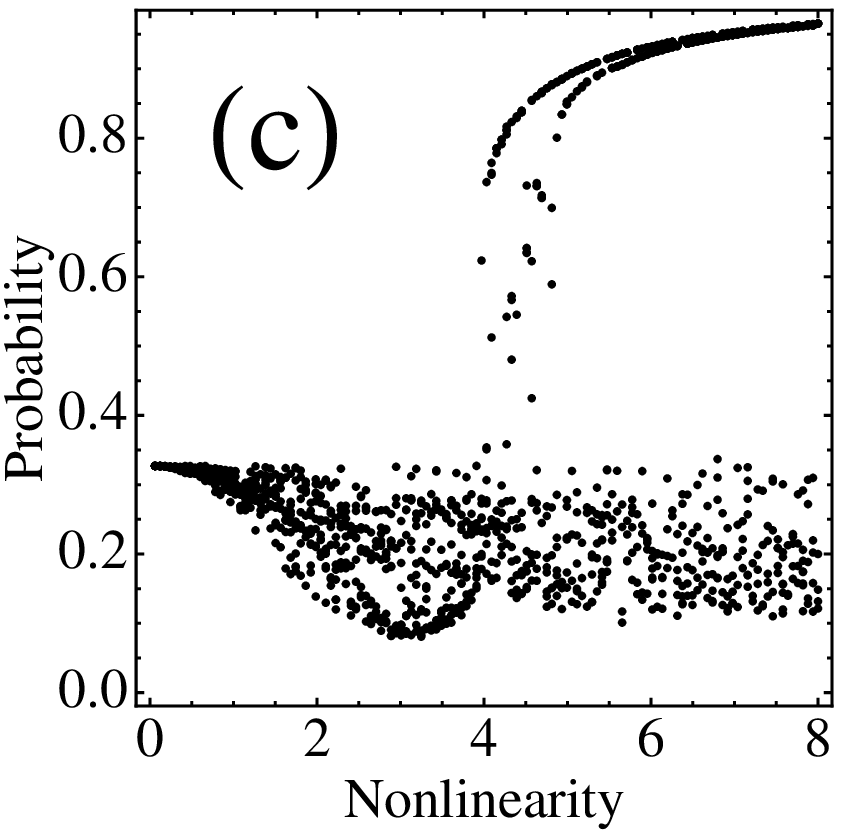}
 \hspace{0.1cm}
 \includegraphics[scale=0.45]{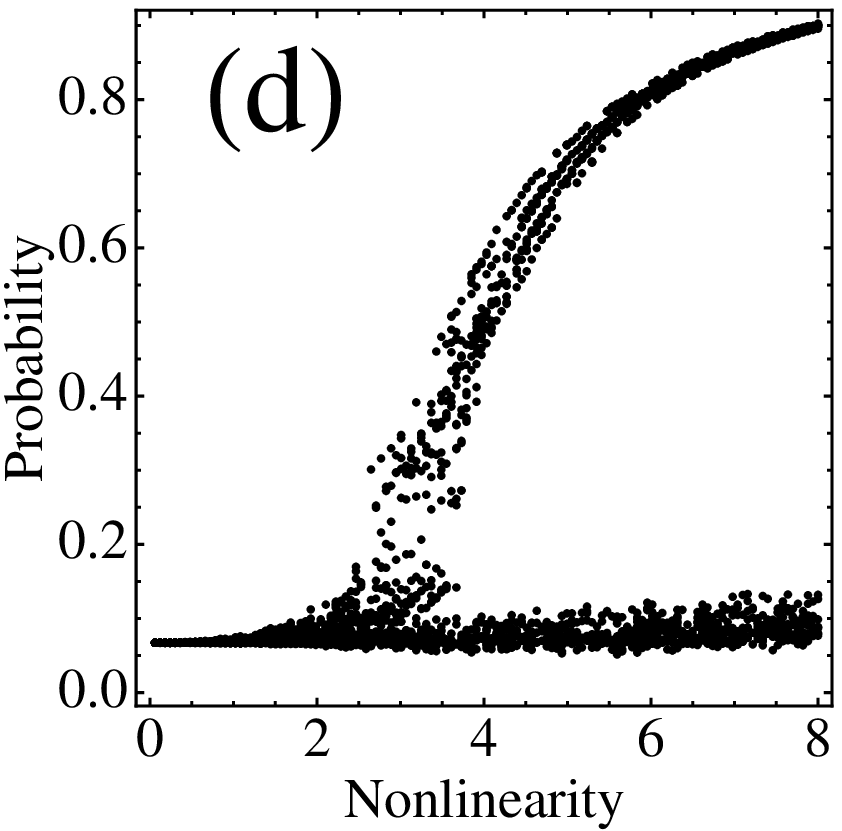}
  \caption{Scatter plot of the time-averaged probability of finding the excitation at the initial site vs the nonlinearity parameter. Top: Lieb lattice with initial condition at (a) site with $2$ nearest neighbors (n.n.) and (b) site with $4$ n.n. Bottom: Stub lattice initial site with $1$ n.n. (c) and $3$ n.n. (d). Twenty realizations are shown.}
  \label{fig3}
\end{figure}
to the selftrapping transition
at high nonlinearities, we also have a large degree of linear selftrapping\cite{tetrahedron}. All this in the  absence of any periodic lattice. This similarity with the rest of the other lattices is no accident and, as we will show below, is due to a common element: degeneracy. As indicated on Fig.1, there are some curves which, at low nonlinearities approach zero (save boundary effects), while others approach a finite value. These curves differ on the excitation site inside the lattice. For instance, in the case of the Lieb lattice, when the initially excited site fall on a site with coordination number equal to two (Fig.2, curve $2$), that site can be thought as belonging to the generic fundamental mode of the Lieb flat (completely degenerate) band. This fundamental mode, in the form of a `ring' (see Fig.1c) has a nonzero overlap with our initial condition, implying that the time evolution of our initial excitation will contain a contribution from the flat mode (as well as contributions from the modes of the extended bands). Since the flat mode has no time evolution, a part of our initial state will not propagate giving rise to a degree of trapping. Now, the situation gets even better: The flat band states (marked with colors in Fig.1) remain as flat modes for the nonlinear problem\cite{dany}. This can be quickly checked for each lattice from the stationary version of Eq.(1). This highly stable nature of the flat band mode, especially for the Lieb lattice, make these systems attractive in optics for the undistorted propagation of signals at long distances. Now, what happens when the initial site does not fall `inside' the fundamental flat band mode? For the Lieb lattice this is the case with $n=4$ nearest neighbors. Here, at low nonlinearity we expect that the pulse will spread all over the lattice, leading to a zero trapped fraction at the origin at long times. Further increase of nonlinearity will ultimately produce the usual selftrapping transition (Fig.2, curve $4$). These conclusions also apply  for the rest of the lattices shown in Fig.1.

For the special case of the simplex, we can compute the linear selftrapping in closed form. We start from an array of $N$ equidistant sites, which is a special case of Eq.(\ref{1}) with $V_{{\bf n},{\bf m}}=V$ and $\epsilon_{\bf n}=0$. We choose one site and place all the excitation there at $t=0$. As we did in the case  of the lattices, we are interested in measuring the amount of excitation remaining on the initial site at large times.
The spectrum of the simplex is well known and consists of $N-1$ degenerate states with eigenvalue $\lambda=-1$: 
$ |1\rangle=(1/\sqrt{2}) (-1,0,\cdots,1)$, $|2\rangle=(1/\sqrt{2}) (-1,0,\cdots,1,0)$, $|3\rangle=(1/\sqrt{2}) (-1,0,\cdots,1,0,0)$,$\cdots$, plus the state $|N\rangle=(1/\sqrt{2}) (1,1,\cdots,1)$, the last one with eigenvalue $N-1$. The time evolution of $C_{0}(t)$, the amplitude of finding the excitation at the initial site, can be expanded in terms of the eigenstates of the simplex. We obtain,
\be
|C_{0}(t)|^2 = {(N-1)^2 + 1\over{N^2}} + 2 \left( {N-1\over{N^2}} \right)
 \cos(N t)
\ee
whose time average gives
\be
\langle |C_{0}(t)|^2 \rangle = {(N-1)^2 + 1\over{N^2}},
\ee
\\
which shows a linear trapping at the initial site. For $n\geq 2$ it increases monotonically with $N$, approaching $1$ at $N\rightarrow \infty$. The reason for this linear trapping resides on the great amount of degeneracy. Something similar occurs in a periodic system with a flat band, where all the states inside the flatband are degenerate.

\section{Random nonlinearity}
\subsection{Selftrapping}
Now we examine the effect of randomness on the nonlinear parameter $\chi$. We take now $\chi_{n}$ from a bivariate distribution $\chi_{{\bf n}}=\chi$ or $0$, with $50\%$-$50\%$ probability, and $\chi=1,2,3,4,5,6$.
Instead of computing the realization-average of (time-averaged) $\langle P_{o}\rangle$ as would be the usual procedure, we decided to do a scatter plot of $\langle P_{0}\rangle$ for each nonlinear disorder realization and combine all of them. Results are shown in Fig.3 for the Lieb and stub lattices (for the diamond lattice results are similar). The main observation is that there are two non-overlapping regimes: One where there is a selftrapping transition at high nonlinearities, and the other where there is no selftrapping transition and the amount of trapping is finite or $O(1/N)$  depending on the position of the initial excited site. There are virtually no intermediate states, independently of the number of random realizations.
 
The explanation for this behavior lies on the observation that, the nonlinearity at the initial site as well as the local environment around this initial site is random. Let us assume that the nonlinearity at the initial site is $\chi$ and that the nearby sites also have nonlinearity $\chi$.  If $\chi$ is large, there will be a tendency to create a seltrapped mode.
This is due to the fact that the nonlinearity is local
causing that a cluster of nonlinear sites can support a nonlinear localized mode, where the nonlinearity of the rest of the sites away from the trapped mode becomes unimportant. The result for this case is that we will have a selftrapping transition around $4 V$. The second
important case is when the initial site has nonlinearity $\chi$ but it is surrounded by a cluster of purely linear sites ($\chi=0$). In this case we are really talking about a nonlinear impurity, whose seltrapping transition happens at a different value, typically between $3 V$ and $4 V$, depending on the position of the initial site and the geometry of the lattice. Another case is when the  initial site and its close vicinity has sites with zero nonlinearity. Here, the excitation will tend to decay quickly and propagate away. Since the amplitude of this wave becomes smaller and smaller as it propagates, nonlinear effects will become less and less important, and the excitation will escape ballistically. Here, $\langle P_{0}\rangle$ will be very small.

Since the conditions at and near the initial site are random, the system will jump from one behavior to the other, from realization to realization. The main behaviors are clear, however, and would have been missed if we have employed a usual average of $\langle P_{0}\rangle$.
The behavior that is not so random occurs at small nonlinearity values. There, the important thing is whether the initial site  has or not  
overlap with the fundamental flat band mode of the lattice. If there is no overlap $\langle P_{0}\rangle$
will be $O(1/N)$, while when there is overlap, $\langle P_{0}\rangle$ will be finite, as explained before.

\subsection{Dynamics}
Having explained the main features of dynamical selftrapping in the presence of nonlinear disorder, we examine now this disorder affects the transport of localized excitations and the dynamical evolution of the fundamental flatband modes of several lattices.  
The case of a one dimensional lattice with nonlinear disorder was treated in Ref.[$40$]. It was found that the presence of nonlinear disorder was only relevant at the beginning of the time evolution. At later times, the mean square displacement quickly converges to a ballistic profile $\sigma(t)^2 \sim t^2$. This was explained as the weakening of nonlinearity as the height of the wave front decreases as it spreads on the lattice. Based on normalization grounds, the amplitude of the propagating wave behaves as $O(1/N)$ rendering the nonlinear term $\chi |C_{{n}}|^2$
unimportant at asymptotically long times.

It is interesting to see if similar behavior also holds for higher dimensional lattices which also possess flat bands. Figure 4 shows the disorder-averaged mean square displacement 
\begin{figure}[t]
 \includegraphics[scale=0.475]{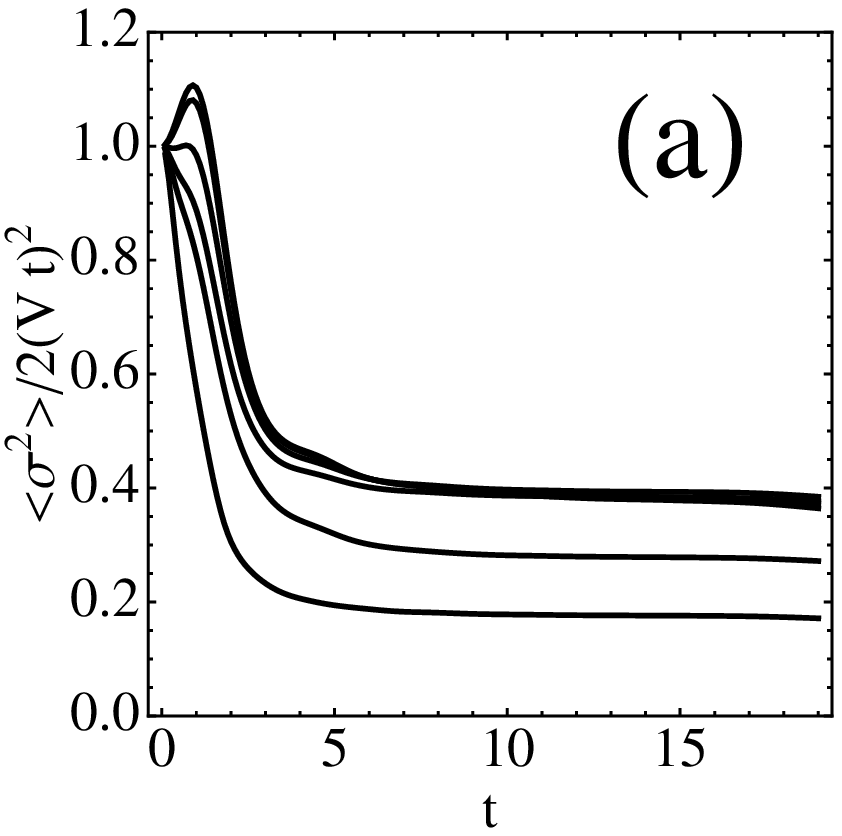}
 \hspace{0.1cm}
 \includegraphics[scale=0.475]{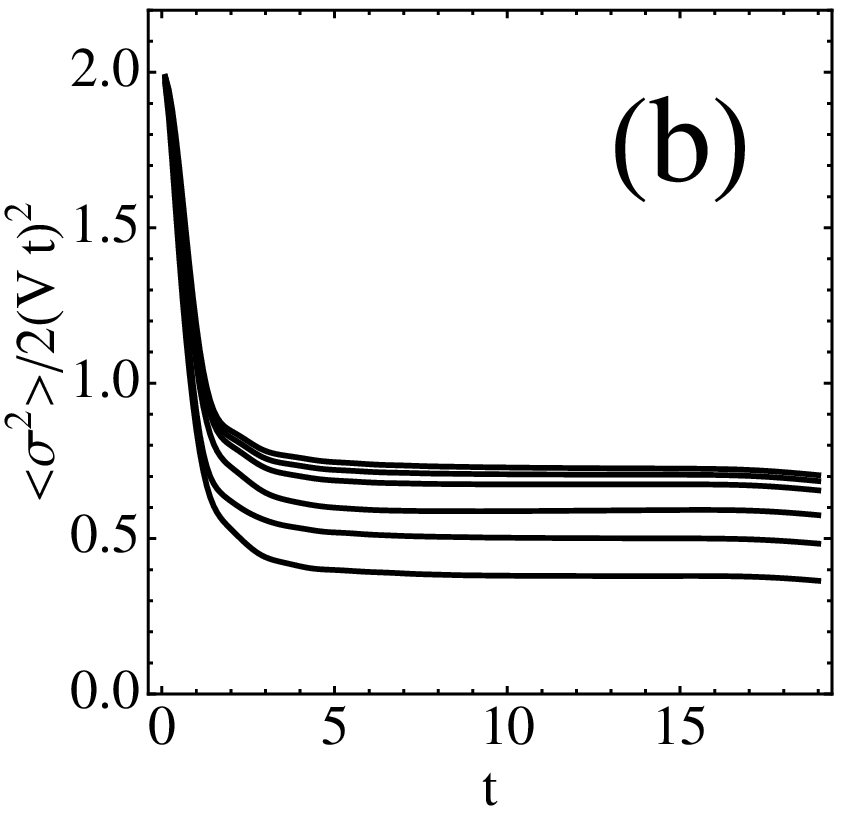}
  \caption{Disorder-averaged mean square displacement for the Lieb lattice, as a function of time, for several nonlinearity values, from $\chi=1$ (highest curve) down to $\chi=6$ (lowest curve). Panels  (a) and (b) refer to an  initial site with and without overlap with fundamental flatband, respectively.}
  \label{fig4}
\end{figure}
for the Lieb lattice, as a function of time, for several different nonlinearity parameter values. In this case, the scatter plots are very 
simple and do not display any internal regimes, and behavior of the system is adequately captured by a realization average.

We consider two cases: One, where the initial site is also part of the fundamental flatband mode (i.e., it has $2$ nearest neighbors), and the other where is not (i.e., it has $4$  nearest neighbors). We have also normalized $\sigma^2(t)$ to the value for the one dimensional ballistic case, $2 (V t)^2$.

We see that after a short interval, all curves converge quickly to the ballistic case, $\langle \sigma^2\rangle \sim t^2$, as in the purely one dimensional case. In   case (a) however, the `speed' $\sigma/Vt$ is lower than in case (b). This is probably due to the fact than in the first case, the initial site has overlap with the fundamental ring mode, giving rise to partial linear trapping. This renormalizes the amount of the wave that can propagate to infinity, giving rise to a lower speed.  Another contribution to the decrease of the speed comes from nonlinearity, as seen already in one dimensional chains\cite{molina_prl}.
\begin{figure}[t]
 \includegraphics[scale=0.45]{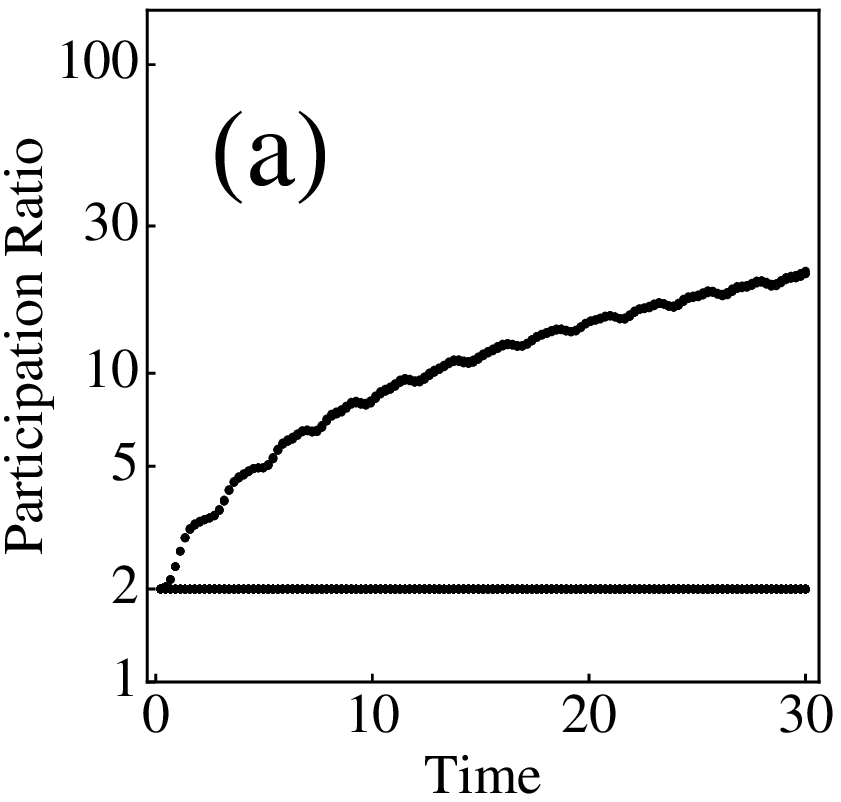}
 \hspace{0.1cm}
 \includegraphics[scale=0.45]{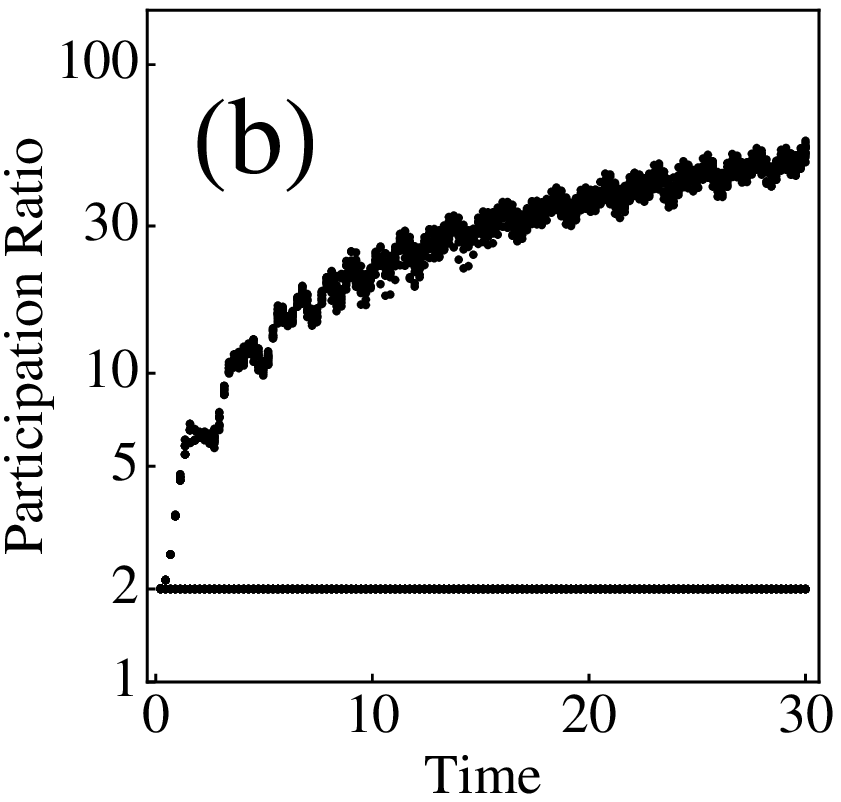}\\
  \includegraphics[scale=0.45]{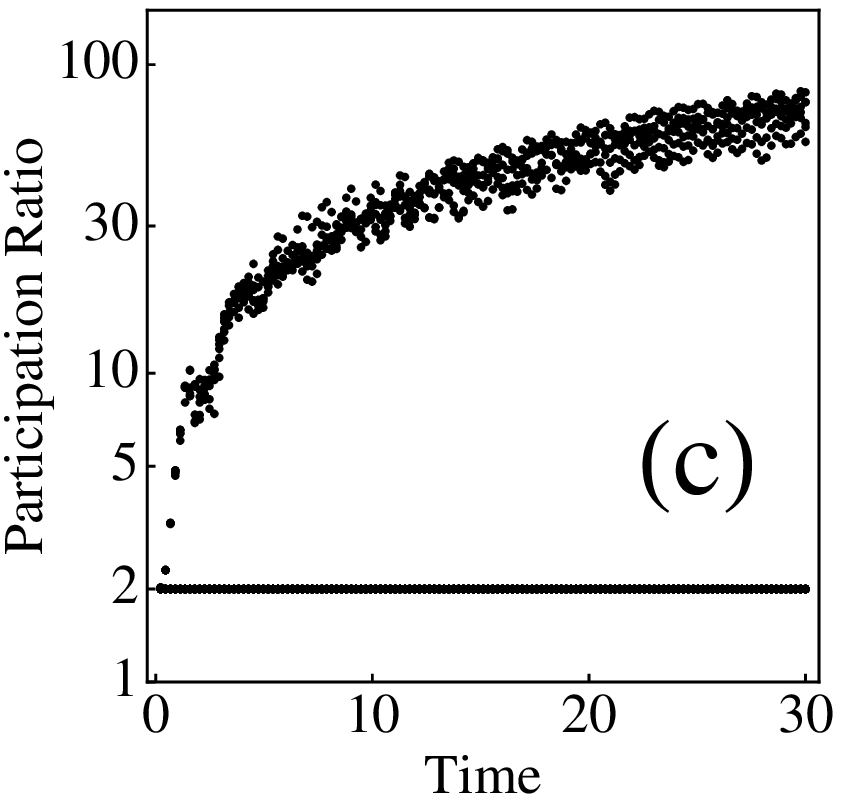}
   \hspace{0.1cm}
   \includegraphics[scale=0.45]{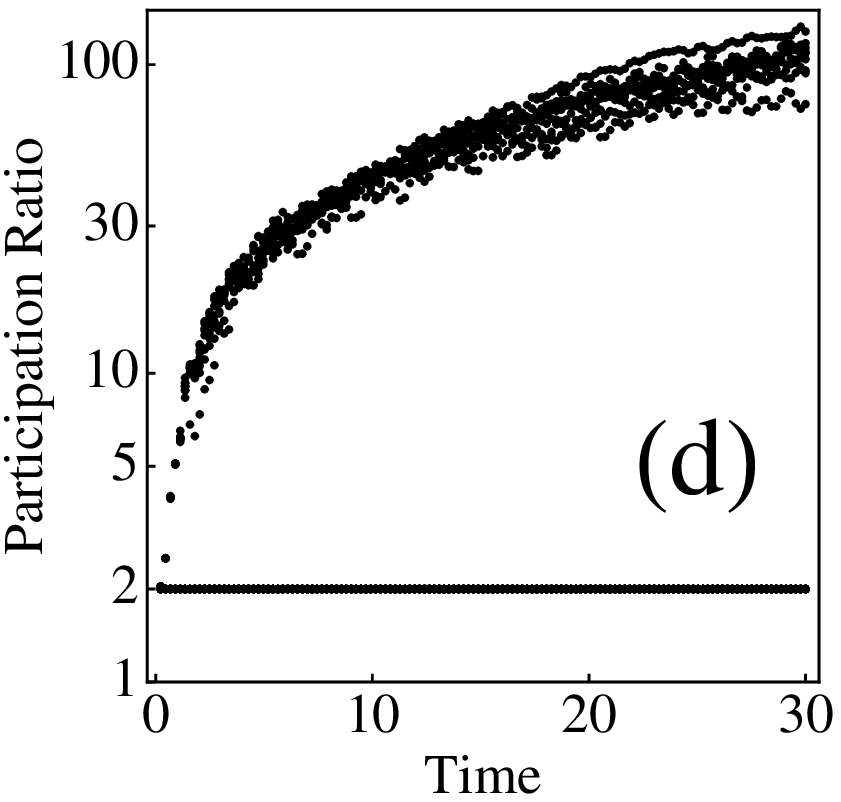}\\
    \includegraphics[scale=0.45]{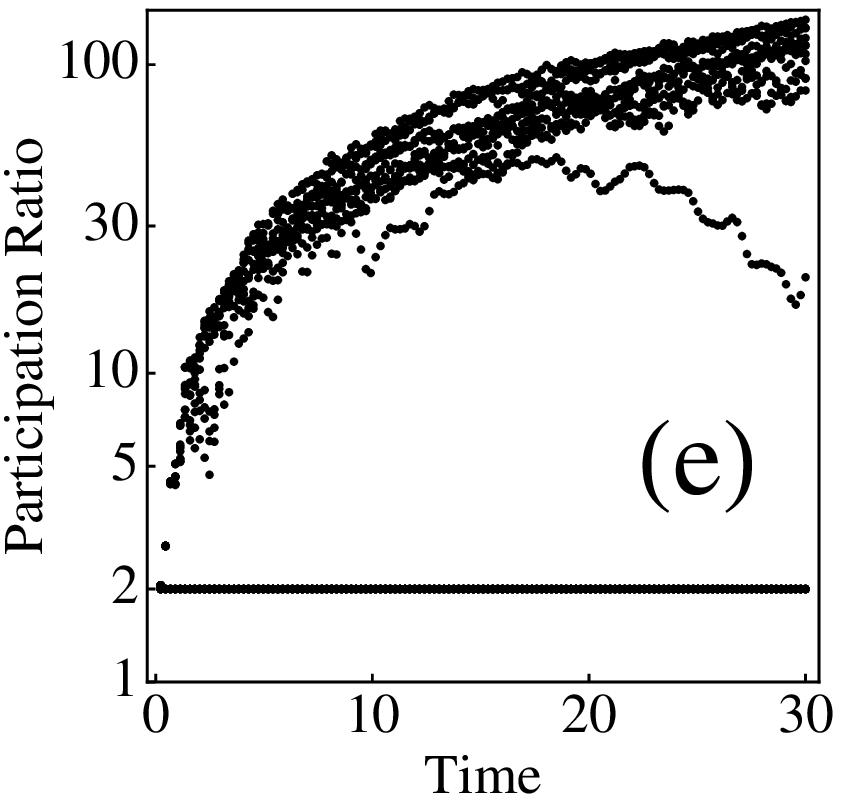}
     \hspace{0.1cm}
     \includegraphics[scale=0.45]{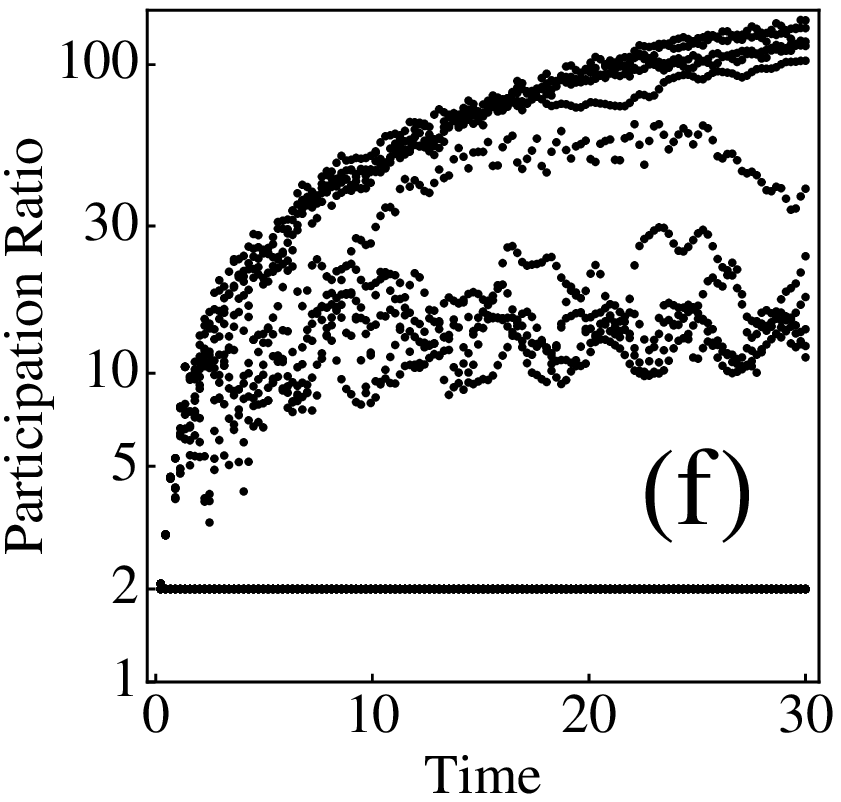}
  \caption{Evolution of the participation ratio $R(t)$ for different nonlinearity strengths: (a) $\chi=1$, (b) $\chi=2$, (c) $\chi=3$, (d) $\chi=4$, (e) $\chi=5$ and (f) $\chi=6$.}
  \label{fig4}
\end{figure}

Finally, we compute the time evolution of the participation ratio $R(t)$ of the fundamental flatband modes in the presence of random nonlinearity. The shape of these modes depend on the particular geometry of the lattice, but they are characterized by a distribution of amplitudes and phases designed to effect an exact phase cancellation that impedes the propagation of the wave beyond a small region, typically consisting on a few sites. Some examples are shown in Fig.1 for the stub, Lieb and diamond lattices.

Figure 5 shows scatter plots for the evolution of the participation ratio $R(t)$ vs time for the diamond lattice, for different nonlinearity disorder $(\chi,0)$ (results for the other two lattices are similar). We immediately notice that the scatter plots reveals two regimes: A spreading one where $R(t)$ increases monotonically with time, and a completely localized regime, where the initial size of the fundamental modes does not change in time, no matter the strength of random nonlinearity. Close scrutiny reveals that $R(t)$ remains constant whenever the nonlinear parameter is the same for both initial sites. Since this happens for half of the time, there is a $50\%$ probability for $R(t)$ to remain at its initial value. This happens for all values of $(\chi,0)$ examined. Notice that this phenomenon would have been missed have we simply computed disorder-averaged curves.

In order to understand this phenomenon of persistence of $R(t)$, we will resort to a simpler model, one that does not have a geometry, but has an almost complete degenerate spectrum: The simplex system. Its main advantage is that we can solve it in closed form. The equations are
\begin{eqnarray}
i (d C_{1}/d t) + V \sum_{m\neq 1} C_{m} + \chi_{1} |C_{1}|^2 C_{1}&=&0\nonumber\\
i (d C_{2}/d t) + V \sum_{m\neq 2} C_{m} + \chi_{2} |C_{2}|^2 C_{2}&=&0\\
\vdots\hspace{2cm}\vdots\hspace{2cm}\vdots           & &\nonumber\\
i (d C_{n}/d t) + V \sum_{m\neq n} C_{m} + \chi_{n} |C_{n}|^2 C_{n}&=&0 \nonumber\\
\vdots\hspace{2cm}\vdots\hspace{2cm}\vdots           & &\nonumber\\
i (d C_{N}/d t) + V \sum_{m\neq N} C_{m} + \chi_{N} |C_{N}|^2 C_{N}&=&0\nonumber\\\label{final}\end{eqnarray}
In the absence of nonlinearity, the (degenerate) fundamental mode consists of a finite amplitude on a given site (let's say site ``1'') with amplitude $A$, and another site (site ``2") with amplitude $-A$, with the rest of the sites having zero amplitude. We will show that, in the presence of nonlinearity, this fundamental mode {\bf is still} an eigenmode of the nonlinear system, provided that $\chi_{1}=\chi_{2}$. To prove that, we begin by evaluating the equations for the system at $t=0$, and conclude that $(d/dt)C_{n}(0)=0$, $n=3,4,\cdots$. Since $C_{n}(0)=0$, $n=3,4,\cdots$, this implies that $C_{n}(t)=0$ for $n=3,4,\cdots$. 
As a consequence, from system (\ref{final}) we have, $C_{1}(t) + C_{2}(t)+\sum_{m\neq 1,2} C_{m}(t)=0$, but since $C_{n}(0)=0, n=3,4,\cdots$ this implies, $C_{1}(t) + C_{2}(t)=0.$
The equations for $C_{1}$ and $C_{2}$ become
\begin{eqnarray}
i (d/dt) C_{1} + V C_{2} + \chi_{1} |C_{1}|^2 C_{1}&=&0\label{mario}\\
i (d/dt) C_{2} + V C_{1} + \chi_{2} |C_{2}|^2 C_{2}&=&0.\label{mario2}
\end{eqnarray}
By means of algebraic manipulations, it is straightforward to prove that $|C_{1}|^2+|C_{2}|^2$ is a constant of motion. Thus,
$|C_{1}(t)|^2+|C_{2}(t)|^2=|C_{1}(0)|^2+|C_{2}(0)|^2=2 A^2$, but since $|C_{1}(t)|^2=|C_{2}(t)|^2$, this implies
\be 
|C_{1}(t)|^2=A^2, \hspace{0.5cm}|C_{2}(t)|^2=A^2\hspace{0.5cm}\forall t
\ee
implying that the participation ratio $R(t)=(A^2+A^2)/(A^4+A^4)=2$, is equal to its initial value. Now, in order for this to be fully consistent, Eqs.(\ref{mario})and (\ref{mario2}) must satisfy the constraint $C_{1}(t) + C_{2}(t)=0$. Inserting this constraint into Eq.(\ref{mario2}) and comparing the result with Eq.(\ref{mario}), we conclude that $\chi_{1}=\chi_{2}$ must be obeyed. Therefore, in the presence of random nonlinearity, whenever $\chi_{1}=\chi_{2}$, the participation ratio will not evolve and will keep its initial value. This is precisely what occurs to the flatband lattices examined here.  In our case, nonlinearity can take the value $\chi$ or $0$ with probability $1/2$, which implies that in fifty percent of the random realizations, $R(t)$ will remain constant, regardless of nonlinearity on sites outside the fundamental mode.

\section{Conclusions}
In this work we have examined the selftrapping and transport properties of localized excitations on various nonlinearly-disordered lattices characterized for having a spectra with flat bands in the linear limit. We found that the presence of high degeneracy due to the presence of flat bands has a strong impact on trapping and transport. The presence of disorder in the nonlinearity causes a fluctuating environment at the initial site, making the system jump between the seltrapped state and the free state. At small nonlinearity, flatbands effects cause linear selftrapping provided the initial site belongs to one of the fundamental flat band modes. The mean square displacement showed a ballistic character at long times, with the nonlinear disorder playing a minor role only. The evolution of the participation ratio of a flatband mode in the presence of this nonlinear disorder showed a strong dependence on the values of nonlinearity on the flatband sites: When their nonlinearity values are different, the participation ratio expand monotonically with time. However, when the flatband sites share the same value of the nonlinearity parameter, the participation ratio remains constant, no matter the values of nonlinearity outside the flatband sites. This happens $50\%$ of the time. We have explained this phenomenon in closed form in terms of the degeneracy of the flatbands, by the aid of the $n$-simplex system. The robustness (fifty percent of the time) of the flatband mode against finite nonlinear disorder, combined with stability against other perturbations\cite{dany}, makes these 
flatband lattices and their accompanying flatband modes  promising candidates for optical applications, as in long-distance diffraction-free transmission of information.

\acknowledgments
This work was supported by Fondecyt Grant 1160177.

\end{document}